\documentclass[11pt,twocolumn,letterpaper]{article}
\usepackage{latexsym,amssymb, amsmath,psfig,subfigure}

\begin{document}

\title{Analytical formulations of Peer-to-Peer Connection Efficiency}

\author{Aaron Harwood\\
\\
Department of Computer Science \& Software Engineering \\
University of Melbourne \\
Victoria, 3010, AUSTRALIA.\\
{\tt aharwood@cs.mu.oz.au}}

\maketitle

\begin{abstract}

Use of Peer-to-Peer (P2P) service networks introduces a new communication paradigm because
peers are both clients and servers and so each peer may provide/request
services to/from other peers. Empirical studies of P2P networks have been
undertaken and reveal useful characteristics. However there is to date little analytical
work to describe P2P networks with respect to their communication paradigm and their
interconnections. This paper provides an analytical formulation and optimisation of \emph{peer connection
efficiency}, in terms of minimising the fraction of wasted connection time. Peer connection
efficiency is analysed for both a uni- and multi-connected peer. Given this fundamental
optimisation, the paper optimises the number of connections that peers should make
use of as a function of network load, in terms of minimising the total queue size that requests
in the P2P network experience. The results of this paper provide a basis for engineering
high performance P2P interconnection networks. The optimisations are useful for reducing bandwidth and power consumption, e.g.
in the case of peers being mobile devices with a limited power supply. Also
these results could be used to determine when a (virtual) circuit should be switched
to support a connection.

\end{abstract}

\section{Introduction}

A number of Peer-to-Peer (P2P) projects have been developed over the last few years and the use of the P2P communication paradigm is becoming widely known~\cite{clark01}.
A peer is a process which can connect to and accept connections from other peers.
Hence peers may provide services and may request services from other peers.
Classification of P2P technology is proposed in~\cite{kant02}.

A large number of P2P specific example applications have arisen: \emph{Napster}\footnote{{\tt www.napster.com}} and Gnutella are well known and widely used file sharing systems\footnote{See {\tt opennap.sourceforge.net} for a long list of other P2P file sharing protocols.}, \emph{Freenet}\footnote{{\tt freenet.sourceforge.com}} is an adaptive P2P network application that permits the publication, replication, and retrieval of data while protecting the anonymity of both authors and readers~\cite{Clarke:2001:FDA}, SETI@home is a well developed system for peers to distribute and process data units and
\emph{PCSCW}~\cite{xia02} provides a P2P based computer supported cooperative work environment. 
Other examples can be found in~\cite{stoicachord,aberer01pgrid, CuencaAcuna2002PlanetPReplication,poon02,yeager02}. 

Using TCP/UDP and sockets provides a standard interface for connecting peers but is
insecure. Secure connections, including peer authentication,
 between peers can be provided using SSL or some
other form of secure communication such as Kerberos. Secure connections take
a longer time to establish but are required for certain kinds of activity such
as resource sharing services where peers should provide only authorised access to local resources, e.g. when providing direct access to the underlying operating system as a generic computer resource.

With classic client/server systems, if the server cannot satisfy a client request then the client may redirect its request to another server by disconnecting and
reconnecting. For example, 
the WWW defines connections between
servers in the form of hyper-links which clients use to search for and access information. Furthermore, until recent developments such as XML the WWW 
servers were not
responsible for the connections, which lead to a significant fraction of their connections becoming invalid. 


The communication patterns that arise via the use of P2P are fundamentally
different to classic client/server. Peers may hop from peer to peer,
disconnecting and reconnecting, in order to access a service.
Also service requests may be passed from peer to peer, following paths
determined by the peers. Unlike a static service network such as DNS, connections between peers dynamically change, e.g. to avoid unnecessary intermediate peers.

A number of empirical studies of P2P networks have been undertaken.
The study by Saroiu, Gummadi at. al.~\cite{sgg} gives detailed
measurements of peer characteristics in the Napster and Gnutella networks. They built crawlers to automatically sample peer characteristics in the live peer networks.
Disappointingly, but not surprisingly, the results show that the behaviour of users in a peer system may be categorised as client-like and server-like.
Approximately 26\% of Gnutella users share no data, 7\% of Gnutella users offer more files than all of the other users combined
 and on average 60-80\% of Napster users share 80-100\% of the files. In general, about 22\% of the participating peers have upstream (from peer) bottleneck bandwidths
 of 100Kbps or less, which makes them unsuitable to provide content and data services. The median session (connected) time is approximately 60 minutes, corresponding to be
 the time taken to connect and download a small number of files. Further studies of P2P networks were reported in~\cite{ripeanu02,markatos02}.

 \subsection{Contribution of this paper}
  
 This paper provides an analytical formulation and optimisation of the fundamental
principles governing a peer connection and service provision. In the first
case a single connection is optimised so as to minimise the total time spent connecting
or connected but unused, given an average arrival rate of requests. In the second
case multiple connections are optimised so as to simultaneously minimise the on time
of each connection while minimising the queue lengths at each connection, given
an average aggregate arrival rate.

The optimisations are useful for reducing bandwidth and power consumption, e.g.
in the case of peers being mobile devices with a limited power supply.  Also
these results could be used to determine when a (virtual) circuit should be switched
to support a connection.

\section{Peer connection efficiency}

\begin{figure}[htbp]
\centerline{\psfig{file=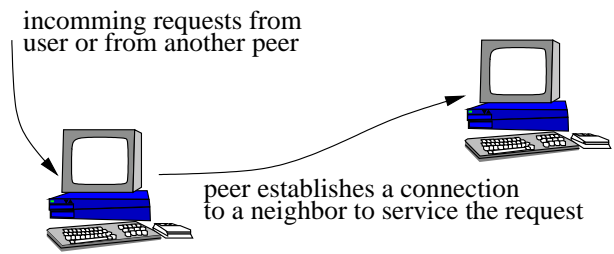,width=\linewidth}}
\caption{Peer-to-Peer connection}
\label{simpleconnection}
\end{figure}

In the following sections the formulation considers a peer which receives requests,
either from a user or from some number of incoming connections, and services these requests
by establishing one or more outgoing connections to other peers, i.e. the peer connects to
another peer in order to service each request as in Figure \ref{simpleconnection}.

\subsection{Single connection}

Let $t_c$ be the time to establish a connection and $t_s$ be the
time taken to complete the service once the connection has been established. Let requests arrive
randomly with an average arrival rate $\lambda$. For a given request, if
the peer is connected (state 1) then the service time is $s_1=t_s$ 
otherwise the peer is not connected 
(state 0) and so the service time is $s_0=t_c+t_s$.

On the arrival of a request let the peer be \emph{not} connected with probability $p$ and connected with
probability $q=1-p$. Then the average service time is
\begin{equation}
E[s]=ps_0+qs_1 = pt_c + t_s.
\label{expected-s}
\end{equation}
The $M/G/1$ queueing system model is applicable with a mean service rate $\mu=1/E[s]$ and utilisation $\rho=\frac{\lambda}{\mu}$.

It follows that, for a fraction $\rho$ of a given time interval, the
peer is either connecting or the connection is being used to provide the service.
For a fraction $1-\rho$ of the same time interval, the peer is either disconnected
or connected but not making use of the connection, the later called idling. One minus the fraction of time that
the peer uses to connect minus the fraction of time that the peer idles
defines the \emph{peer connection efficiency}: 
\begin{equation}
\eta(p;\lambda,t_c,t_s)=1-\rho\frac{pt_c}{pt_c+t_s} - (1-\rho)(1-p).
\label{connection-efficiency}
\end{equation}
Conversely, not using the
connection when it is not needed is considered efficient, as is not having 
to establish a connection every time the service is requested.

As an intuitive example,
consider a hard drive system that has two states: spinning and
sleeping. If the hard drive is sleeping when a request arrives, then it must
first spin up, taking time $t_c$, before servicing the request, taking an
additional time $t_s$. If the hard drive is already spinning, then the request
is serviced in time $t_s$ only. Clearly the system aims to minimise the fraction
of time spent spinning up and idling.

This goal is trivial for $\lambda\rightarrow 0$  by setting $p=1$ and when $\lambda \geq \frac{1}{t_s}$ by setting $p=0$. In the first case the peer is not required
and in the second  case such a situation
is not good because the service becomes unstable as the request queue grows
to infinity. 
Maximising $\eta$ with respect to $p$ yields an optimal value:
\begin{equation}\nonumber
\begin{split}
\frac{d(1-\eta)}{dp}&=2\lambda pt_c+\lambda t_s -1\\
\Rightarrow p^*&=\frac{1-\lambda t_s}{2\lambda t_c}; 0 < \lambda \leq \frac{1}{t_s}.
\end{split}
\end{equation}
However for small $\lambda$, $p^*$ becomes $>1$ so
\begin{equation}
p^*=\begin{cases}
1 & \lambda \leq a,\\
\frac{1-\lambda t_s}{2\lambda t_c} & \frac{1}{t_s}\geq \lambda > a,\\
0& \lambda > \frac{1}{t_s}.\end{cases}
\label{optimalp0}
\end{equation}
where $a=\frac{1}{2t_c+t_s}$.

\begin{figure*}[htbp]
\begin{center}
\subfigure[theory]{\label{theory}\psfig{file=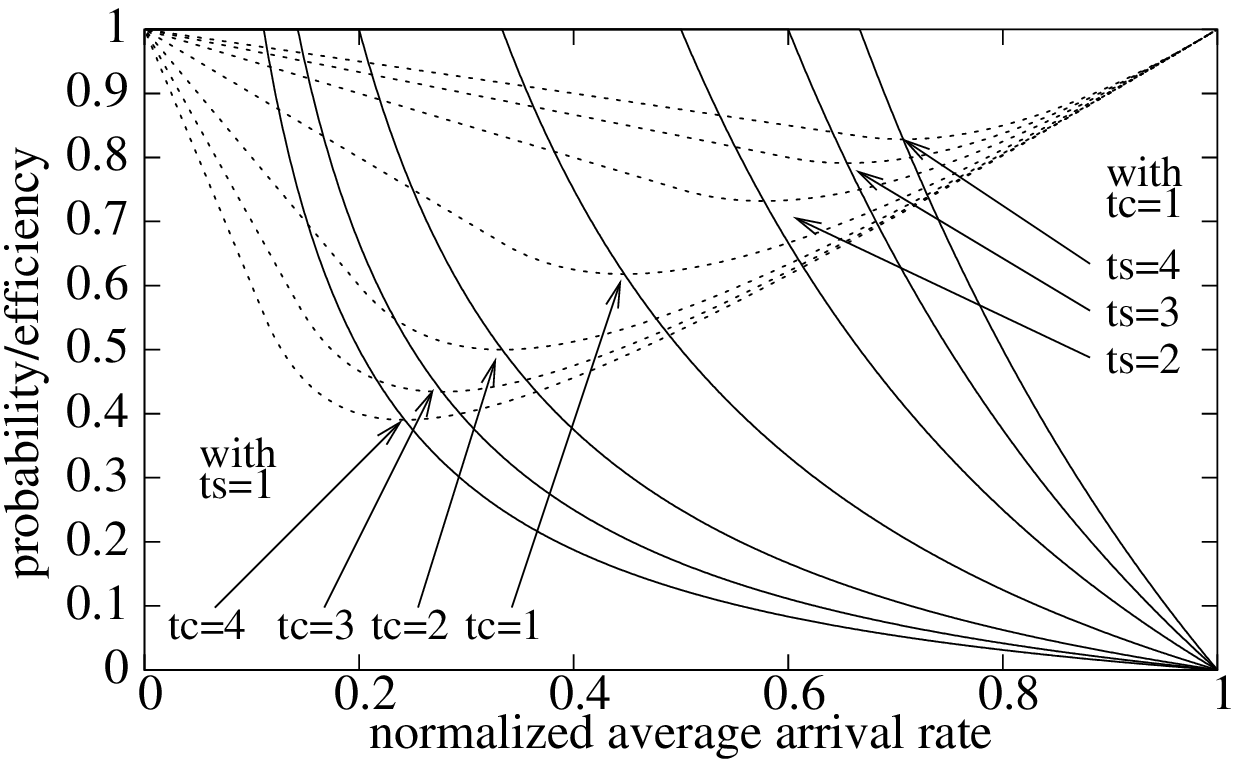,width=3.2in}}
\subfigure[simulation]{\label{exp}\psfig{file=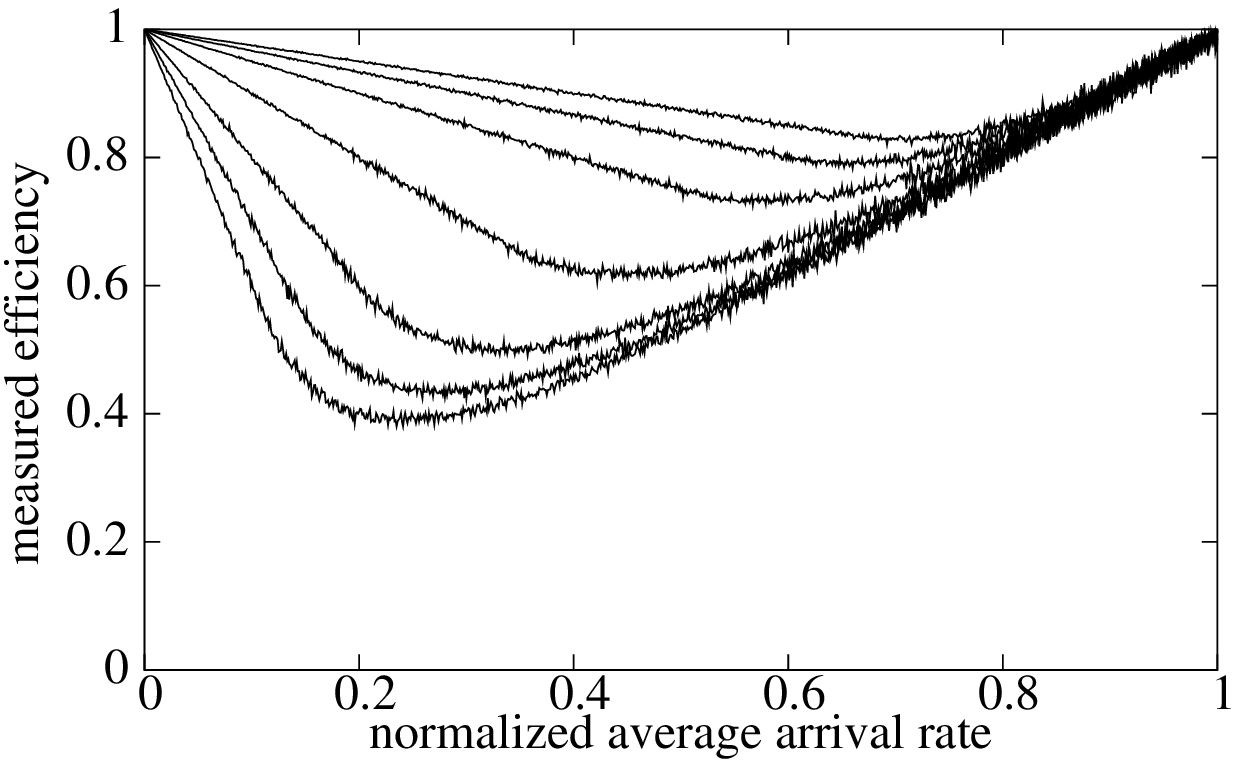,width=3.2in}}
\end{center}
\caption{Probability of disconnecting that gives optimal efficiency versus
normalised average arrival rate for various values of $t_c$ and $t_s$.}
\label{theory-exp}
\end{figure*}

Figure \ref{theory} shows $p$ (series that tend to 0) and $\eta$ (series that
tend to 1) as a function of normalised average arrival rate\footnote{$\lambda$ takes the values $(0,\frac{1}{ts})$.} for various
values of $t_c$ and $t_s$. The series for $t_c=1,2,3,4$ are with $t_s=1$
and the series for $t_s=2,3,4$ are with $t_c=1$. Arrows point to the intersection
of associated series. Figure \ref{exp} shows the results of a simulation (given
in Section \ref{simalg}) that
uses $p^*$ from Equation \ref{optimalp0}. 

\subsection{Multiple connections}

Each peer may open a number of simultaneous connections in order to satisfy
service requests. Since a peer may satisfy a request transparently by simply
passing the request to another peer, any peer may satisfy any type of request.
With this assumption, a peer may take an incoming request and pass it to
any other peer for service. A network of file mirroring peers gives an example
where this assumption holds, every file is available on every peer.

In this case, let requests arrive at the peer with an average arrival
rate of $\Lambda$ and let the peer maintain $d$ (outgoing) connections, with
each connection receiving an average
$\lambda_i$, $i=0,1,2,\dotsc,d-1$, requests per second where
\[
\sum \lambda_i = \Lambda.
\]
In other words, the peer
forwards a request to connection $i$ with probability $\lambda_i/\Lambda$.
Let $\mathbf{\lambda}=(\lambda_0,\lambda_1,\dotsc,\lambda_{d-1})$.
Note that the peer must observe the bounds $0 \leq \lambda_i < \frac{1}{t_s}$
to ensure that the queue lengths associated with each connection are finite.

In the previous analysis the probability $p$ that a connection is disconnected
after servicing a request was computed such that the wasted time, being
the sum of connecting time and idle time, was minimised. Continuing, let
$p_{i}$ be this probability for connection $i$. 
From Equation \ref{connection-efficiency} let 
\[
\xi_i=p_{i}(1-\rho_i)
\]
be called the \emph{off-time} of connection $i$ where $\rho_i$ is the utilisation
of connection $i$. Let $t_{c,i}=t_c$ and $t_{s,i}=t_s$ be invariant over all connections. When $p_{i}$ is set to minimise the fraction of wasted time
for a given $\lambda_i$ then it maximises the fraction of time available
for both servicing the requests and for residing in the off state. 

When $\xi_i=0$ the connection $i$ is on for all time and conversely when $\xi_i=1$
the connection $i$ is off for all time. For a given $\lambda_i$, the length of
the request queue that connection $i$ is servicing\footnote{Using Pollaczek's formula.} is
\[
L_i=\frac{\rho_i^2}{1-\rho_i}  \frac{1+C_{s,i}^2}{2}
\]
where from Equation \ref{expected-s}
\[
C_{s,i}^2=\frac{Var[s,i]}{E[s,i]^2}=\frac{t_c^2p_{i}(1-p_{i})}{(p_{i}t_c+t_s)^2}
\]
is the squared coefficient of variation for the $M/G/1$ queueing system.
The quantity $\xi_i/L_i$ is the fraction of time the connection is off per
request. It follows that
\begin{equation}
OPR\big(\mathbf{\lambda};p_{i}^*, t_c, t_s\big)=\frac{\sum_{i=0}^{d-1} \xi_i}{\sum_{i=0}^{d-1} L_i}
\label{offtimeperrequest}
\end{equation}
is the average off time per request for the peer.
Clearly the objective of the peer is to maximise $OPR$
for a constant $\Lambda$.

\begin{figure*}[htbp]
\begin{center}
\subfigure[$d=2, t_c=1, t_s=1$]{\label{lambda-2}\psfig{file=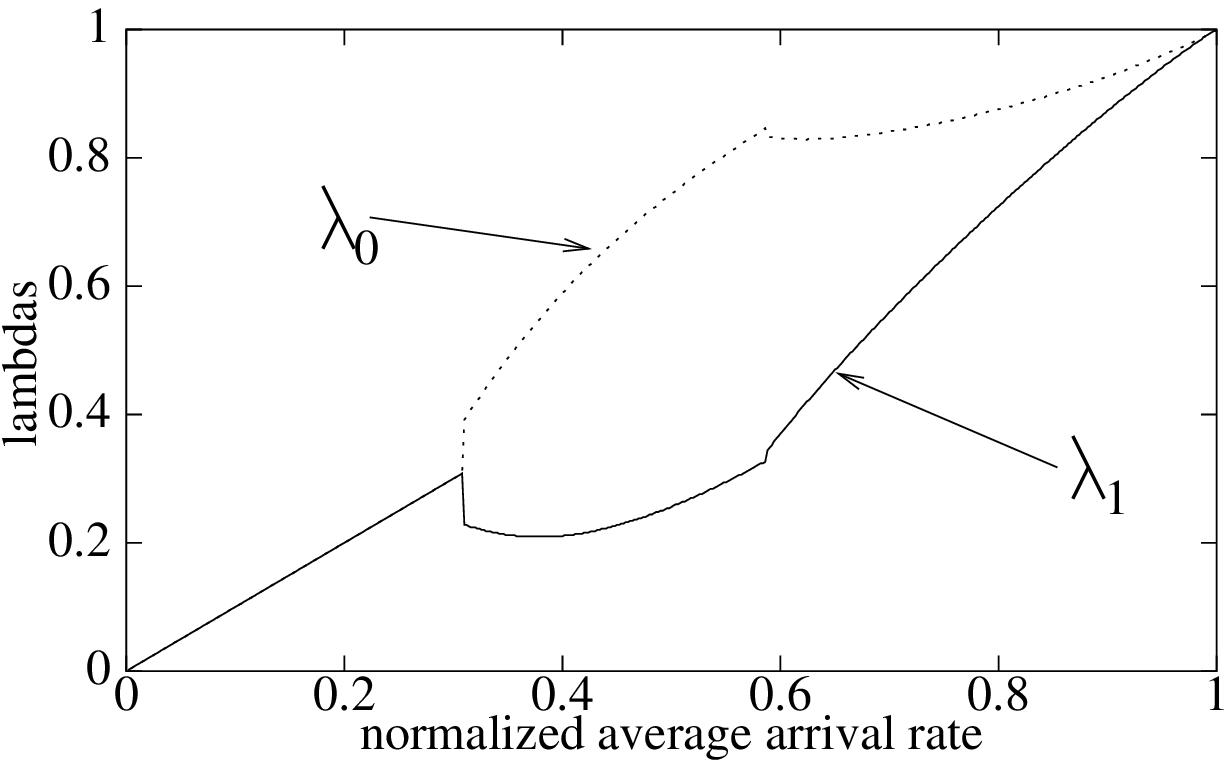,width=3.2in}}
\subfigure[$d=3, t_c=1, t_s=1$]{\label{lambda-3}\psfig{file=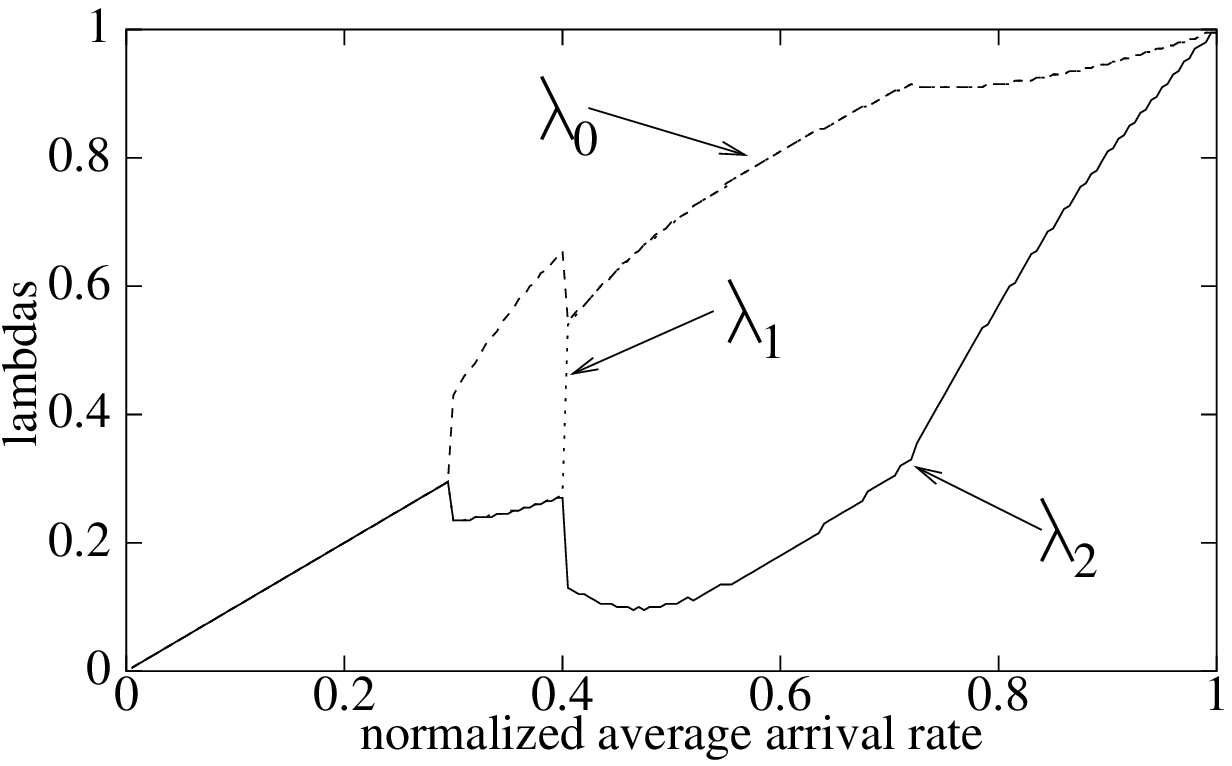,width=3.2in}}
\subfigure[$d=5, t_c=1, t_s=1$]{\label{lambda-5}\psfig{file=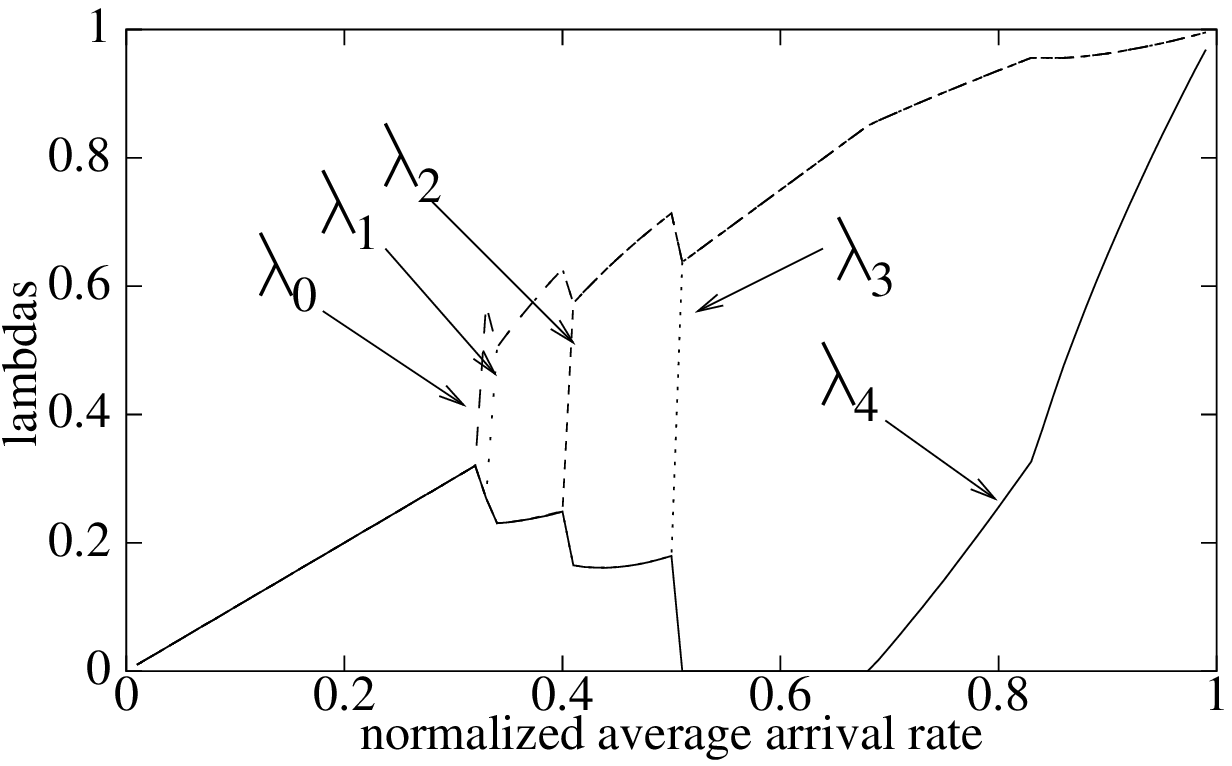,width=3.2in}}
\subfigure[$d=10, t_c=1, t_s=1$]{\label{lambda-10}\psfig{file=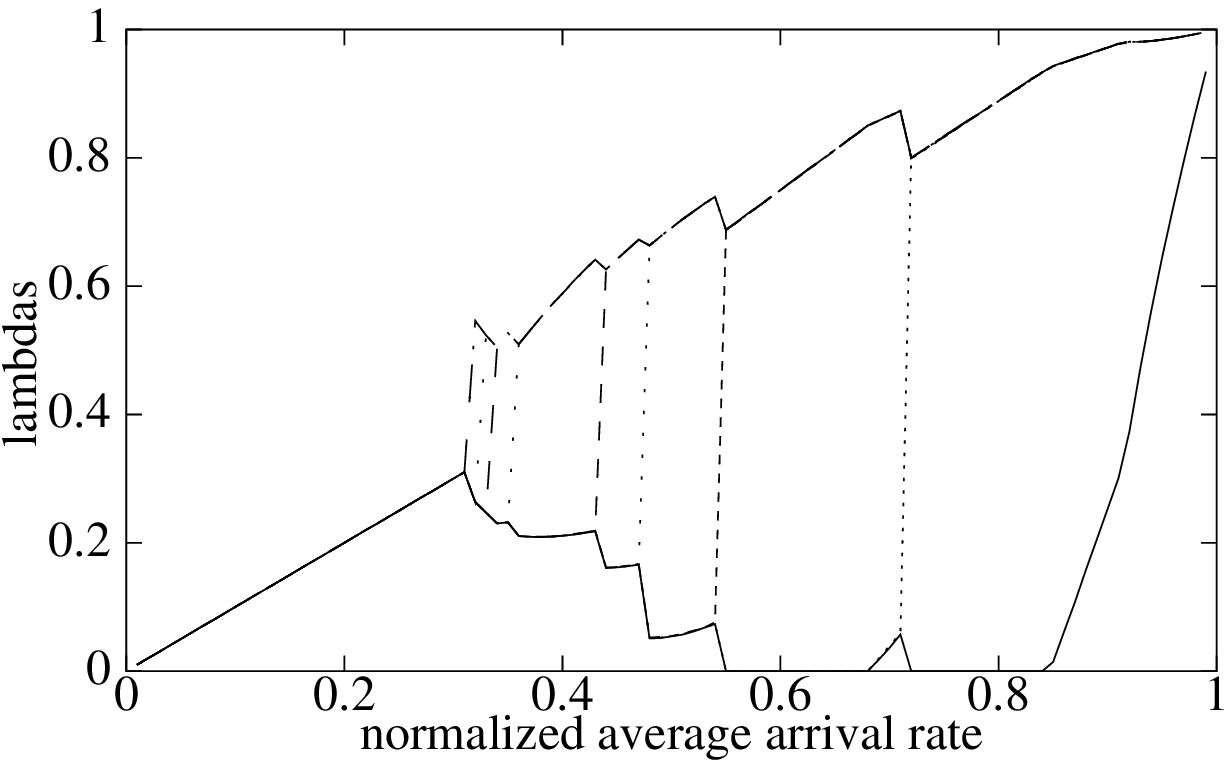,width=3.2in}}
\subfigure[$d=5, t_c=5, t_s=1$]{\label{lambda-5.5.1}\psfig{file=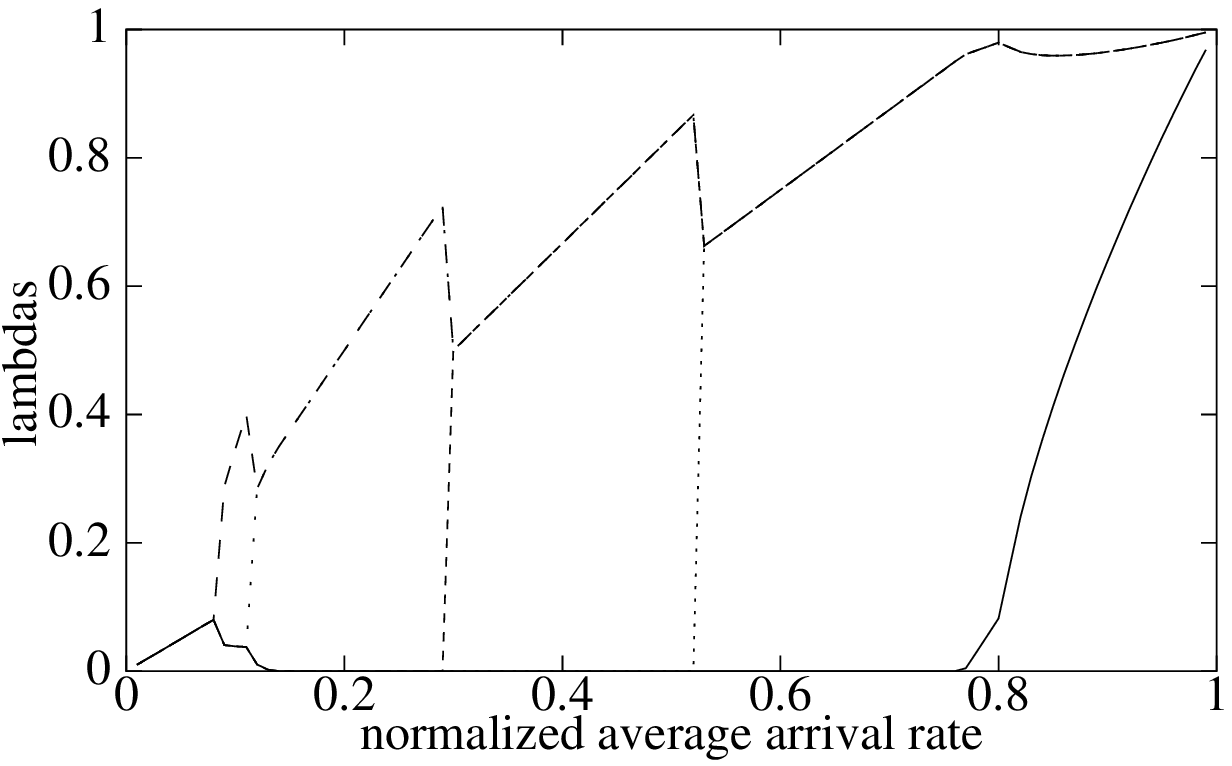,width=3.2in}}
\subfigure[$d=5, t_c=1, t_s=5$]{\label{lambda-5.1.5}\psfig{file=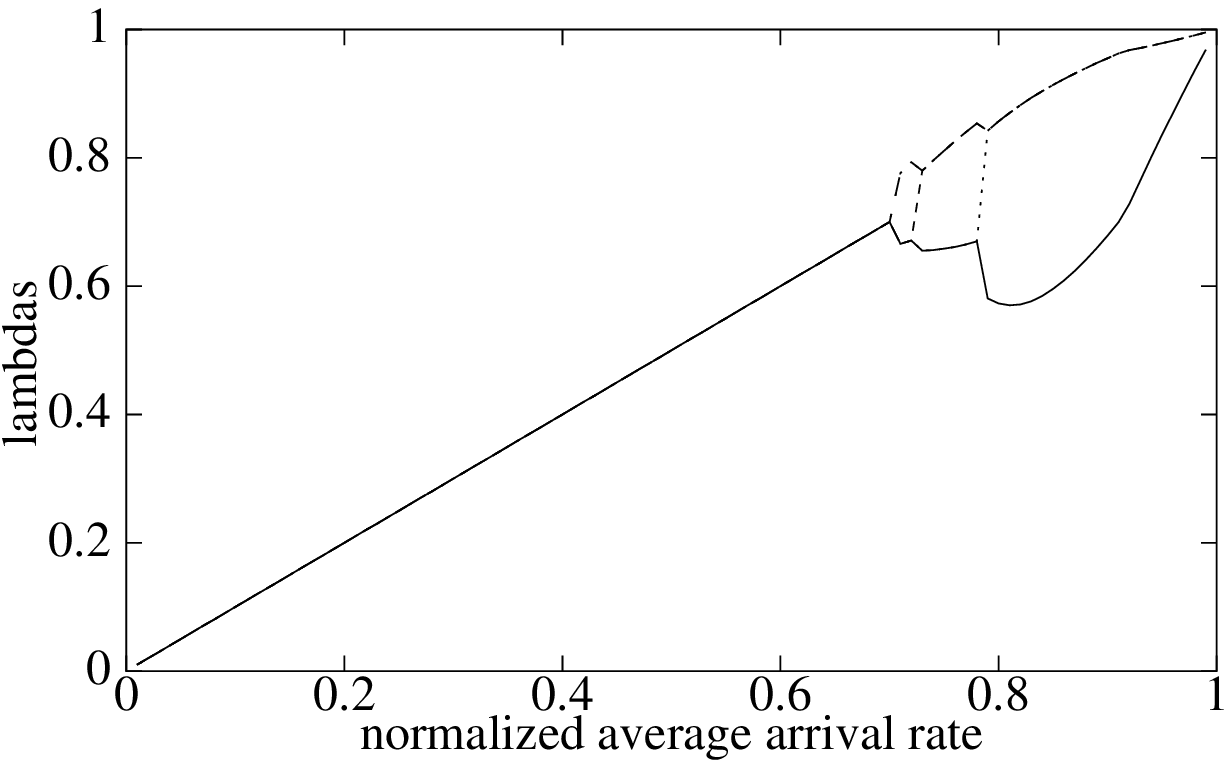,width=3.2in}}
\end{center}
\caption{Values for $\lambda_i$ that maximise $OPR$ over
$d$ possible connections versus $\Lambda$.}
\label{lambda-solutions}
\end{figure*}

Numerical solutions, using simulated annealing, that maximise Equation \ref{offtimeperrequest} for the cases when $d=2,3,5,10$ are shown in Figure \ref{lambda-solutions} with $\Lambda=(0,\frac{d}{t_s})$. In
Figure \ref{lambda-2}, $\lambda_0=\lambda_1$ for small $\Lambda$ and similarly
in Figures \ref{lambda-3}, \ref{lambda-5} and \ref{lambda-10} (however $\lambda$'s are
difficult to distinguish and are not shown for this last case).
In Figure \ref{lambda-5.1.5} $\lambda_0$ and $\lambda_1$ are equal.

From Figure \ref{lambda-solutions} it is seen that the peer should partition
the set of connections into a set with a high utilisation and a set with low utilisation in order to achieve the objective. The vertical lines show when a $\lambda_i$ switches from a low
value up to a high value. For certain regions where the solution involves
a $\lambda_i=0$ it is evident that less connections
is better. 

Clearly if $t_c > t_s$ then optimal conditions are reached by loading a small
number of connections and when $t_c < t_s$ the connections may be equally loaded most of the time. Mirrored hard drives, which have spin up time much greater than the time to service
a request would operate optimally according to Figure \ref{lambda-5.5.1}. Peer connections tend
to be established faster than the time to service the request so they would operate
optimally according to Figure \ref{lambda-5.1.5}.

\subsection{Servicing a fraction of requests}

\begin{figure}[htbp]
\centerline{\psfig{file=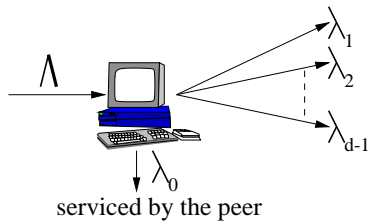,width=0.6\linewidth}}
\caption{A peer servicing $\lambda_0$ of the $\Lambda$ requests and forwarding those remaining.}
\label{sopr}
\end{figure}

\begin{figure*}[htbp]
\begin{center}
\subfigure[$d=3, t_c=1, t_s=1$]{\label{lambda3.s.1.1}\psfig{file=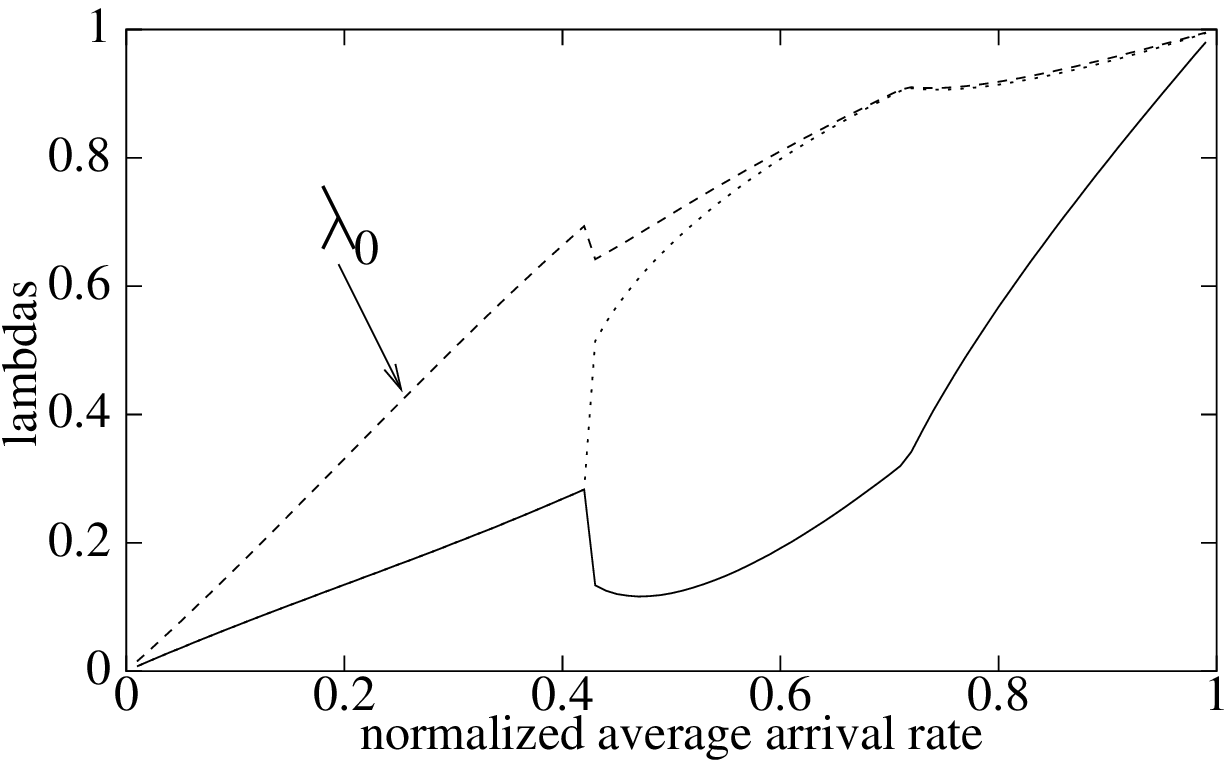,width=3.2in}}
\subfigure[$d=5, t_c=1, t_s=1$]{\label{lambda5.s.1.1}\psfig{file=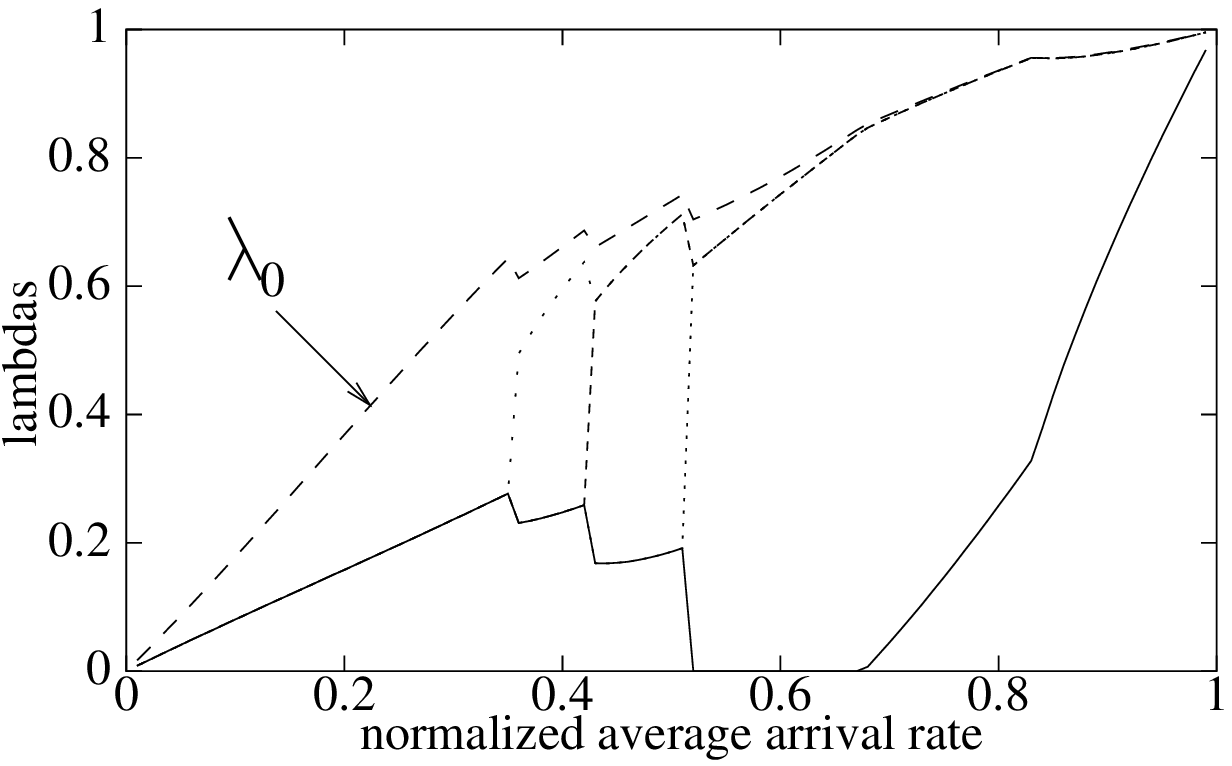,width=3.2in}}
\end{center}
\caption{Values for $\lambda_i$ that maximise $SOPR$ over
$d$ possible connections versus $\Lambda$.}
\label{lambda-SOPR}
\end{figure*}

The peer is expected to service some fraction of the total $\Lambda$ requests. Let $\lambda_0$ be the chosen
rate of requests that the peer services as depicted in Figure \ref{sopr}. In this case Equation \ref{offtimeperrequest} is modified to
obtain a service off time per request:
\begin{multline}
SOPR\big(\mathbf{\lambda};p_{i}^*, t_c, t_s\big)=\\ \frac{\sum_{i=1}^{d-1} \xi_i}{L_0+\sum_{i=1}^{d-1} L_i},
\label{serviceofftimeperrequest}
\end{multline}
where $L_0=\rho_0^2/(1-\rho_0)$ and $\rho_0=\lambda_0t_s$. There is no $t_c$ for requests
serviced by the peer and no wasted time since the peer is on all the time.
Figures \ref{lambda3.s.1.1} and \ref{lambda5.s.1.1} show how $SOPR$ compares to $OPR$ for
the case when $d=3$ and $d=5$ respectively.

\subsection{Expected number of peers traversed}

So far the time $t_s$ has been considered the time taken to service a request
over an established connection. Consider $t_s$ to be the time taken to transmit the
request. In this case, the request is transmitted from peer to peer until it is finally
serviced. The probability of being serviced at a peer is
\[
p_s=\frac{\lambda_0}{\Lambda}
\]
and the average number of peers that a request passes through (including
the last one) is then $k=1/p_s$.

The average queue length experienced by a request as it passes through a
peer is
\[
\hat{L_{q}} = \sum_{i=1}^{d-1}L_{q,i}\frac{\lambda_i}{\Lambda-\lambda_0}
\]
and so the total length of queues experienced by a request in the P2P network
is
\[
\sum_{i=1}^{\infty} p_s(1-p_s)^{i-1}\left(\hat{L_{q}}(i-1) + L_0\right)
\]
where $L_0$ is defined for $SOPR$. Average queue length increases in direct proportion
to the number of connections that each peer maintains. This suggests that having
a small number of connections is better. However, because the connections
are not bandwidth constrained, having more connections allows a larger load.
Computing optimal connection efficiency given bandwidth constraints required
dynamic programming techniques and is left for further research.


\section{Conclusion}

Deciding if and when to disconnect from a server in order to minimise the
total wasted connecting plus idle connection time with respect to the arrival
of requests is an interesting problem. This maximises the \emph{connection
efficiency}. This paper derived a probability $p$
such that, for a given request arrival rate $\lambda$, if the client (blindly) disconnects with probability $p$ after servicing a request then the connection efficiency is maximised. 
In this case the client has no knowledge of the queue size. 
It is seen that, for a connection establishment time $t_c$ and a request
service time $t_s$, the worst efficiency occurs when the request rate is about
$\frac{1}{2t_c+t_s}$.
Better efficiency could
be obtained by considering the queue size (whether $>0$ or not) and other statistical information that describes
the arrival of requests. This aspect is left for further study.

In a P2P network, peers can connect (like clients) to other peers (like servers) 
in order to service incoming requests. Using a broad assumption, peers can relay requests to other peers so that
any peer can transparently service any kind of request. As each connection from a peer has
an associated connection efficiency, which is a function of request rate, it follows
that the average connection efficiency over the peer's connections is dependent on the
proportion of requests forwarded over each connection. To avoid solutions that
give unbounded queue sizes the paper proposed a quantity called \emph{off time per request} and uses simulated
annealing techniques to find an assignment of request rates over 
a given number of connections that maximises this quantity, thus minimising
wasted time.

Furthermore, the paper considered each peer to be servicing some fraction of requests
that it receives, rather than forwarding these requests to other peers. In this case
a similar quantity called \emph{service off time per request} is introduced and
maximised subject to similar conditions. 

It was seen that the forwarding of requests to peers should be divided into two general
classes: a highly loaded class and a lightly loaded class. This intuitively follows from observing
that the least efficient operating load for a connection, as a function of load, is
between 0 and maximum load and that a lightly loaded connection is less efficient
that a heavily loaded connection. For more
than a few connections it is seen that some connections are better left inactive
until required to handle the load.

Future research will study the delay (wait time) rather than queue length. Although delay and
queue length are related as
\[
W_q = \frac{L_q}{\lambda}
\]
the solutions for minimum service off time per request are significantly different.
Thus, the system must choose to minimise $W_q$ or $L_q$ or a combination
of both. Also, analytical solutions to the multidimensional minimisation problem
have not be derived. The problem is a piecewise non-linear minimisation problem with
constraints. The method of simulated annealing required a significantly large number
of restarts in order to achieve the global minimum. 

The study of peer connection efficiency is an interesting area of P2P research that
describes fundamental P2P modes of operation using rigorous analytical theory.
There is substantial grounds for further analysis and potential for
new discoveries.

\appendix

\section{Simulation algorithm}
\label{simalg}

The following simulation was executed with $T=5000$ arrivals to obtain
the results of Figure \ref{exp}.
\begin{enumerate}
\item given $t_c$ and $t_s$ and a uniform random variable $X$
between $0$ and $1$,
\item for each $\lambda$
\begin{enumerate}
\item compute $p$ according to Equation \ref{optimalp0}
\item $Connected=false$
\item $Waste=0.0$
\item $CurrentTime=0.0$
\item $ArrivalsTime=0.0$
\item for $T$ number of arrivals do
\begin{enumerate}
\item $Arrival=(-1/\lambda)\log(1-X)$
\item $ArrivalsTime=ArrivalsTime+Arrival$
\item $TimeDifference=ArrivalsTime-CurrentTime$
\item if $TimeDifference<0$ then set $TimeDifference=0$
\item if $Connected=true$ then
\begin{enumerate}
\item set $Waste=Waste+TimeDifference$
\item set $CurrentTime=CurrentTime+TimeDifference+t_s$ 
\end{enumerate}
\item ELSE
\begin{enumerate}
\item set $Waste=Waste+t_c$
\item set $CurrentTime=CurrentTime+t_s+t_c$
\end{enumerate}
\item set $Connected=false$ with probability $p$ and $=true$ with probability $1-p$
\end{enumerate}
\item output $\eta(\lambda)= 1-Waste/CurrentTime$
\end{enumerate}
\end{enumerate}

\bibliographystyle{plain}
\bibliography{p2p,paper,survey}

\end{document}